# The Role of Parsimonious Models in Addressing Mobility Challenges

Marc Barthelemy[1,2,*]

[1] Université Paris-Saclay, CNRS, CEA, Institut de Physique Théorique, Gif-sur-Yvette, France.
[2] Centre d'Analyse et de Mathématique Sociales CAMS, UMR 8557 CNRS-EHESS, Ecole des Hautes Etudes en Sciences Sociales, Paris, France
* Email: `marc.barthelemy@ipht.fr`

## Abstract

Mobility is a complex phenomenon encompassing diverse transportation modes, infrastructure elements, and human behaviors. Tackling the persistent challenges of congestion, pollution, and accessibility requires a range of modeling approaches to optimize these systems. While AI offers transformative potential, it should not be the sole solution. Parsimonious models remain crucial in generating innovative concepts and tools, and fostering collaborative efforts among researchers, policymakers, and industry stakeholders.

## Challenges for sustainable mobility

As cities evolve and their populations increases, the question of how people move within (and between) urban environments becomes increasingly critical. With the growing accessibility of data [1] (in particular mobile phone data [2]), numerous models for human mobility have been developed [3], and more recently machine learning techniques [4]. The various modeling approaches are employed to address the challenges of congestion, pollution, and inefficiency in transportation, and play a crucial role in understanding the dynamics of transportation systems and designing effective strategies to improve their performance. We note that the future of modeling mobility is however not just about predicting traffic patterns or optimizing public transportation; it is about envisioning ~~holistic,~~ sustainable, and inclusive solutions that cater to the diverse needs of urban dwellers, a much more difficult goal to achieve. Sustainable mobility faces several challenges ranging from infrastructure and technological innovations, multimodal integration, environmental planning, to more social sciences related questions such as affordability, equity and accessibility, and behavioral changes. The list of these challenges makes it clear that the solutions cannot be purely technological and will need an interdisciplinary effort, ranging from civil engineering to social sciences.

Many papers discuss the future of urban mobility and possible scenarios [5, 6] for transportation modes likely to be offered in the coming years. In particular, car-dominated urban mobility seems to face an uncertain future, and autonomous and electric vehicles might not provide a solution that can compete with mass transit [6], at least from the point of view of delays due to congestion [7]. It is therefore crucial to explore how urban mobility could evolve, using simulations that incorporate new modes of transport, mobility options, and multimodality. At a more "physical" level, sustainable mobility relies on urban planning strategies that prioritize walkability, mixed land use, and transit-oriented development. However, implementing these principles often conflicts with existing zoning regulations, development patterns, and automobile-centric infrastructure designs. Also, it is essential to develop the infrastructure to support sustainable transportation modes such as public transit, cycling lanes, charging stations for electric vehicles. There are however some complex challenges: electric vehicles rely on energy grids powered by fossil fuels in some regions, and cycling infrastructure may have environmental impacts during their construction. Balancing the environmental benefits of sustainable mobility with potential drawbacks is essential and need to be taken into account in modeling these problems.

Also, modeling social aspects and human responses to changes will be essential to address these questions, which presents obvious challenges. The authors in [8] explore various significant factors, including the influence of spatial cognition on human mobility, the formation of new mental maps affecting movement behaviors, the adaptation of humans to emerging transportation modes, and the impact of algorithms on our mobility patterns. Encouraging individuals to adopt sustainable travel behaviors requires overcoming entrenched habits and attitudes. Education, incentives, and behavioral nudges can help promote mode shifts towards more sustainable options (and more equity), but changing societal norms and cultural perceptions of mobility will take time, and modelling these effects is a very difficult task. In this context, it can be noted that AI techniques for opinion mining can be very useful for assessing citizen's opinion (see for example [9]).

Enumerating these diverse challenges highlights the intricate nature of mobility, emphasizing that its modeling is far from straightforward. From a practical standpoint, addressing these challenges requires collaboration among governments, businesses, civil society organizations, and individuals, further complicating the search for solutions. By tackling infrastructure gaps, promoting behavioral change, fostering innovation, and prioritizing equity, sustainable mobility can become a reality for communities around the world, but it will be very likely difficult to achieve [10, 11]. It is therefore crucial to identify the tools that will help us to achieve these goals. Models, predictions, and simulations play a pivotal role in this contex. Specifically, we will argue here that parsimonious models provide a valuable perspective that should not be overlooked or dismissed.

## Predicting vs. understanding

The scientific study of mobility and transport is at least one hundred years old and it is impossible to discuss the whole history of this field here. Historically, earlier discoveries were often driven by the need for simplicity, leading to the development of parsimonious models (also called 'descriptive'). Here and throughout the subsequent discussion, when we mention a parsimonious model, we are referring to a set of assumptions regarding the main mechanisms and ingredients, and calculations conducted within this framework establish relations between various quantities. What qualifies as parsimonious can be open to subjective judgment, and we can find a large range of models with a number of parameters going from zero or one to much more. A famous saying attributed to Von Neumann [12] says that "with four parameters I can fit an elephant, and with five I can make him wiggle his trunk" which gives a sense of what is meant by parsimonious modeling for a physicist. Ideally, a parsimonious model should incorporate a minimal set of parameters and mechanisms, enabling it to account for a broad range of empirical observations. This balance between parameter count and data validation is what renders parsimonious models truly valuable. Notably, the gravity model, introduced in the 1930s and still employed today for predicting the traffic flow between two areas, identified key parameters governing human mobility [13, 14, 15]. Additional approaches emerged, including discrete choice models [16], or intervening opportunities [17], with subsequent efforts focused on enhancing these approaches. For instance, the radiation model, proposed in 2012, exemplifies a parsimonious (yet with predictive capabilities) model [18].

More generally, over the past century, a wide array of approaches has been suggested, ranging from purely empirical methods to highly theoretical frameworks, some of which have not been validated with data. Despite the diversity of these studies, two broad categories of models can be identified based on their primary objectives: understanding or predicting. Predictive models are designed to forecast future trends and outcomes, utilizing numerical techniques such as agent-based models, microsimulations (see for example [19] and references therein), and, more recently, AI techniques. In contrast, models aimed at understanding seek to reveal the underlying mechanisms and dynamics that drive mobility patterns. These models are usually parsimonious, often relying on analytical approaches but occasionally employing numerical methods, such as simulations on very simple systems (Figure 1). In addition, we can in general delineate two sub-categories within parsimonious approaches: those primarily aimed at comprehending a phenomenon and highlighting dominant mechanisms, and those that, while simplifying reality, possess the capability to provide quantitative predictions validated by empirical data.

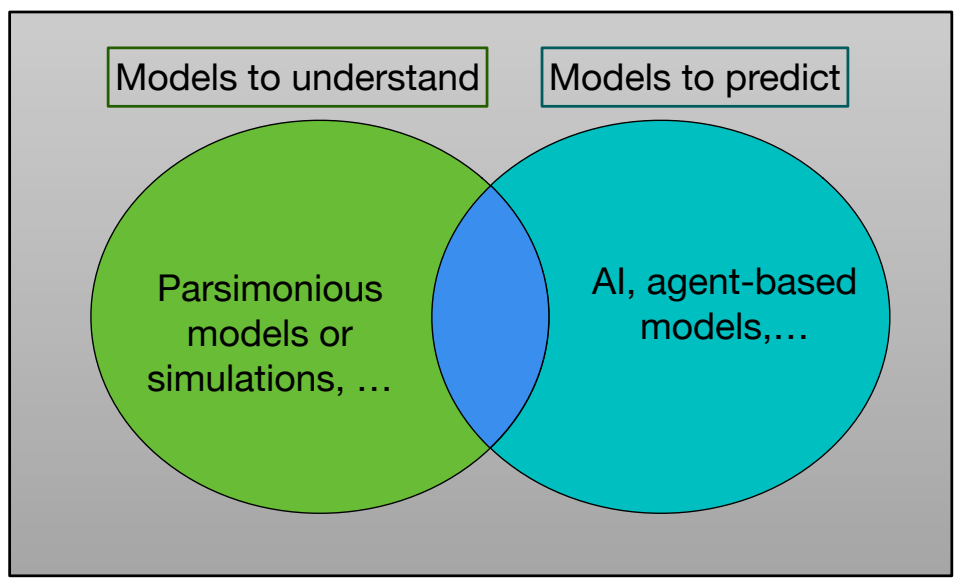

**Figure 1**: Categories of models: understanding versus predicting. We distinguish between models primarily aimed at understanding a phenomenon (in green) and those focused on predicting future outcomes (in cyan). For predictive models, the main approaches include agent-based models and, more recently, AI techniques. In contrast, models designed to enhance our understanding typically rely on parsimonious models. Notably, some parsimonious models also possess predictive capabilities, as indicated by the blue intersection in the diagram.

Here, we explore the interplay between predictive and models to understand (mainly represented by parsimonious models), highlighting their respective strengths and roles in modeling transportation systems.

**Predictive models: artificial intelligence**

While predictive models include all sorts of techniques (such as agent-based and other simulation techniques), we will focus here on the very promising use of Artificial Intelligence (AI). AI has the potential to address major societal problems, including sustainability [20], and plays also a pivotal role in shaping the future of cities [21, 22], human mobility and transportation systems [23]. This include predictive maintenance [24], autonomous vehicles management [25], traffic management [26], ride-sharing and mobility services, public transportation optimization [27], personalized navigation, etc. We cannot here discuss all these applications but for the example of traffic management, AI algorithms (such as reinforcement learning for example [28]) optimize traffic flow by adjusting signal timings, rerouting vehicles to avoid congested areas, reducing delays, and improving travel times. The impact of this sort of approach will certainly continue to grow in the near future. In the case of public transportation optimization [29], AI can enhance public transit systems by predicting demand (on a short time scale so far), optimizing schedules, and improving route planning, which benefits commuters and reduces environmental impact. Basically, AI algorithms analyze historical data, real-time traffic conditions and passenger demand patterns. From this analysis, these algorithms optimize bus and train schedules in order to minimize waiting times for passengers, and predictive models help anticipate peak hours and allocate resources accordingly, ensuring timely services. AI can also help in network design such as optimizing bus stop locations for example [30]. AI also powers self-driving cars [31], enabling them to perceive their surroundings, make decisions, and navigate without human intervention. These vehicles promise safer roads, reduced congestion, and improved efficiency. However, it is important to note that embracing emerging technologies such as autonomous vehicles, shared mobility services, and mobility-as-a-service platforms presents challenges related to data privacy, cybersecurity, and regulatory frameworks (see for example the review about privacy issues in human mobility studies [32]. Other applications include: road condition monitoring thanks to AI-powered sensors, and can help transportation agencies prioritize maintenance and repairs, ensuring safer roads for commuters; traffic incident detection thanks to video feeds analysis (from surveillance cameras), that detect accidents, breakdowns, or roadblocks promptly, such that authorities can respond faster, minimizing disruptions and ensuring smoother traffic flow; and others such as pedestrian detection, driver monitoring, smart parking management, automated license plate recognition, ride-sharing and mobility as a service, AI-driven navigation apps, etc.

Basically, all these predictive models harness historical data and statistical algorithms to forecast future mobility patterns, offering invaluable insights for planning and decision-making. In particular, machine learning techniques such as regression analysis, time series forecasting, and neural networks excel at extrapolating trends and identifying patterns within complex datasets. By leveraging predictive models, it is clear that cities can anticipate traffic congestion, optimize public transit schedules, and allocate resources more effectively.

While AI tools offer numerous benefits in transportation, they also come with several drawbacks and challenges, and the most important ones are the following. First, AI algorithms rely heavily on data for training and decision-making. However, transportation data can be incomplete, inaccurate, or outdated, leading to biased or unreliable predictions and decisions [33]. Second, AI algorithms can perpetuate or amplify biases present in the data used for training [34]. For example, if the input transportation data reflects existing inequalities or discriminatory practices, AI tools may inadvertently reinforce these biases in decision-making processes. Third, many AI algorithms, such as deep learning models, are often regarded as "black boxes" due to their complex structures and internal workings. Although this is not always regarded as a bad thing [35], this lack of interpretability makes it challenging to understand how AI tools arrive at their decisions, hindering transparency and accountability in transportation systems.

## Parsimonious models: critical parameters and dominant mechanisms

Parsimonious models, on the other hand, focus on understanding the underlying dynamics and causal relationships driving mobility behavior, in sharp contrast with AI tools. These models often employ mathematical modelling, statistical techniques, network analysis, and simulation methods to dissect complex urban systems. By uncovering factors such as land use patterns, socioeconomic demographics, and spatial interactions, parsimonious models provide a nuanced understanding of why certain mobility patterns emerge and evolve over time. These models are often a caricature of reality, sometimes simplifying it to the extreme by retaining only a very small number of parameters and mechanisms. They usually take into account crucial ingredients from complexity science such as short-term and long-term dynamics and multiple interactions between agents in the system [36]. In transportation, these models aim to describe and analyze the current state or behavior of transportation systems without necessarily making predictions or prescribing actions, but help transportation planners, engineers, and policymakers understand existing patterns, identify trends, and assess the performance of transportation systems. Some examples of these models include traffic flow, travel demand, mode choice, or environmental models, illustrating the wide range of their use.

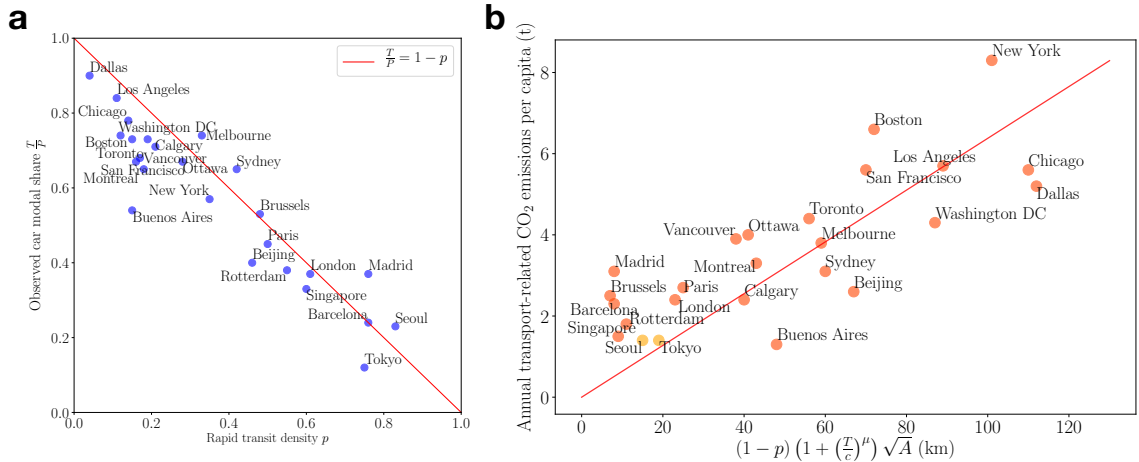

**Figure 2**: Modal share and $CO_2$ emissions in cities. (a) Modal share of cars versus the probability $p$ to have access to a rapid transit station, tested for various cities in the world. The red line is the prediction of a simple model [37], without any tunable parameter. (b) Prediction for the transport-related emitted $CO_2$ from the model described in [37]. Even if fluctuations are not negligible, this model allows to identify the critical parameters for $CO_2$ emissions: the surface area of the city, the access to mass rapid transit and congestion effects. Figures from [37].

There is therefore a very large number of such models in the scientific literature and we cannot list them all. For the sake of concreteness, we discuss this idea of parsimonious models on two simple examples [37, 38] that illustrate their capacity to help us understand a phenomenon and sometimes also to make predictions. In the first example [37], we considered a simplified model of mode choice between car and mass rapid transit for the journey-to-work: an individual has a probability $p$ to be in the vicinity of a mass rapid transit station and in this case has to compare the generalized cost of car versus mass transit (and with probability $1 - p$, the individual takes his car). This model leads to the very simple result that the share of car traffic $T/P$ (car users over the population) is equal to the proportion $1 - p$ of individuals that do not have access to the mass rapid transit. Such a result can be tested on various cities and despite its simplicity proved to be very accurate (see Figure 2, left). Furthermore, this model allows us to compute the $CO_2$ emitted per capita (which is roughly proportional to the gasoline consumption per capita) by cars in cities, and the calculation leads to a product of three factors:

$$Q_{CO_2} \sim (1-p)\sqrt{A}(1+\tau) \qquad (1)$$

where $A$ is the surface area of the city, and $\tau$ is the additional delay due to congestion (Figure 2, right). This model indicates that the renowned finding of Newman and Kenworthy [39], which suggests that gasoline consumption per capita is a "simple" decreasing function of urban density, is likely to be incorrect. The result (1) suggests that if we want to mitigate the $CO_2$ emission (or similarly the gasoline consumption), an important ingredient is to increase the public transport density or the access to it. This simple model thus supports quantitatively the transit-oriented development idea [40] and allows us to both predict some quantities and to understand what is the dominant mechanism (here the accessibility to a mass rapid transit station). However, this generic model should be applied cautiously before drawing broad policy conclusions and must be tailored to specific contexts.

In contrast, there are models that may not offer predictions for individual cases but nonetheless enable the identification of dominant mechanisms and enhance our understanding of a phenomenon. For instance, investigating empirically the sources of population growth in cities leads to a stochastic equation [38] describing the temporal changes of its population $P(t)$:

$$\frac{dP}{dt} = \eta P + DP^\beta \zeta. \qquad (2)$$

where $\eta$ is a gaussian noise, $\zeta$ a Levy noise (and $D$, $\beta$ are positive constant). An important point to emphasize is that this equation reveals that urban population growth originates primarily from two separate sources: stochastic growth resulting from the balance between births and deaths (corresponding to the first term of the r.h.s. with a gaussian noise), and a second source (term with the Levy noise) arising from inter-urban migrations. Even if this equation cannot predict what will happen for a given city, an interesting result is that this inter-urban migration term described by a Levy noise has the property to exhibit very large fluctuations that correspond to migration "shocks". This shows that the fate of a city is not written in stone but can be influenced by policies and governance. This is here an example of a parsimonious model that allows us to identify the main mechanism of a phenomenon, even if it cannot predict the future of a specific case.

There are obviously many other instances, but the aim here is to showcase the strengths of a parsimonious model: if corroborated by data, it elucidates the primary mechanisms and identifies the crucial parameters—an endeavor that proves challenging for AI tools.

## Discussion

While predictive and parsimonious models serve distinct purposes, and even if the distinction between them is sometimes blurred, their integration holds probably the key to comprehensive urban mobility analysis. Predictive models can inform short-term decision-making and tactical interventions, such as optimizing traffic signal timings or rerouting public transit routes. Meanwhile, parsimonious models offer a deeper understanding of long-term trends, facilitating strategic planning initiatives like urban redevelopment or transportation infrastructure investments. Advanced technologies enable predictive models to incorporate real-time data feeds, enhancing their accuracy and responsiveness to dynamic urban environments. Similarly, parsimonious models benefit from access to large-scale, granular datasets, allowing for more nuanced insights into emerging mobility phenomena. The synergy between predictive and parsimonious models fosters feedback loops that drive continuous learning and improvement. Predictive models generate hypotheses about future mobility trends, which can be supported by parsimonious modeling techniques. Conversely, parsimonious models uncover hidden patterns and anomalies that inform the development of more robust predictive algorithms, creating a virtuous cycle of knowledge generation and refinement. Even if the limits of AI in our understanding of cities and mobility need to be explored [41], the integration of predictive and parsimonious models will hopefully emerge as a fundamental pillar of urban mobility planning and governance.

**Acknowledgements**

Not applicable

**Competing Interests**

The author declares no competing interests

**Data availability statement**

Not applicable